\documentclass[prb,twocolumn,superscriptaddress]{revtex4}
\usepackage{graphicx}
\usepackage{bm}
\usepackage{amsmath}
\usepackage{indentfirst}

\newcommand{\ph}[1]{\phantom{#1}}

\begin{document}

\title{Spontaneous polarization and piezoelectricity in boron nitride
nanotubes}

\author{S. M. Nakhmanson}
\email{nakhmans@nemo.physics.ncsu.edu}
\homepage{http://nemo.physics.ncsu.edu/~nakhmans/}
\affiliation{Department of Physics, North Carolina State
University, Raleigh, NC 27695}
\author{A. Calzolari}
\affiliation{INFM --- National Research Center on nanoStructures and bioSystems at Surfaces (S3) 
and Dipartimento di Fisica, Universit\`a di Modena e Reggio E., Modena, Italy}
\author{V. Meunier}
\affiliation{Department of Physics, North Carolina State
University, Raleigh, NC 27695}
\affiliation{Center for Computational
Sciences (CCS) and Computer Science and Mathematics Division, Oak
Ridge National Laboratory, Oak Ridge, TN 37830}
\author{J. Bernholc}
%\affiliation{Department of Physics, North Carolina State
%University, Raleigh, NC 27695}

\author{M. Buongiorno Nardelli}
\email{mbnardelli@ncsu.edu}
\affiliation{Department of Physics, North Carolina State University,
Raleigh, NC 27695}
\affiliation{Center for Computational
Sciences (CCS) and Computer Science and Mathematics Division, Oak
Ridge National Laboratory, Oak Ridge, TN 37830}

\date{\today}

\begin{abstract}
\textit{Ab initio} calculations of the spontaneous polarization and
piezoelectric properties of boron nitride nanotubes show that they
are excellent piezoelectric systems with response values larger
than those of piezoelectric polymers. The intrinsic chiral
symmetry of the nanotubes induces an exact cancellation of the total
spontaneous polarization in ideal, isolated nanotubes of arbitrary
indices. Breaking of this symmetry by inter-tube interaction or
elastic deformations induces spontaneous polarization comparable to
those of wurtzite semiconductors.

\end{abstract}


\maketitle

\section{introduction}

Piezo- and pyroelectric materials for modern technological
applications should display an excellent
piezoelectric response, combined with high mechanical stability and
low environmental impact. Existing materials, which can be broadly
divided into the families of ceramics and polymers, can only partially
fulfill the above requirements.  Lead zir\-co\-nate titanate (PZT)
ceramics, for example, are strong piezo- and pyroelectrics \cite{PZT,PT} 
%with spontaneous polarizations of 0.3--0.7 C/m$^2$. 
but, unfortunately, they are
also brittle, heavy and toxic. On the other hand, polymers like 
polyvinylidene fluoride (PVDF) are lightweight, flexible and virtually
inert, but their polar properties are an order of magnitude weaker than
those of PZT.\cite{PVDF} In this paper, we examine spontaneous polarization
and piezoelectricity in boron nitride nanotubes (BNNTs) in order to
estimate their potential usefulness in various pyro- and piezoelectric
device applications, and to understand the interplay between symmetry
and polarization in nanotubular systems.

BNNTs, broadly investigated since their initial pre\-dic\-ti\-on
\cite{prediction} and succeeding discovery,\cite{discovery} are
already well known for their excellent mechanical properties.
\cite{crespi2000} However, unlike carbon nanotubes, most of BN 
structures are non-centrosymmetric and polar, which might suggest the
existence of non-zero spontaneous polarization fields. Recently, these
properties have been partially explored by Mele and Kr\'al, using a model 
electronic Hamiltonian.\cite{mele02} They predicted that BNNTs are
piezo- and pyroelectric, with the direction of the spontaneous electric field  
that changes with the index of the tubes. 
The \textit{ab initio} calculations presented in this paper
provide a much fuller description and show that BNNT systems are
indeed excellent lightweight piezoelectrics, with comparable or better piezoelectric
responce and superior mechanical properties than in piezoelectric polymers. 
However, contrary to the conclusions of Ref.~\onlinecite{mele02}, 
our combined Berry phase and Wannier 
function (WF) analysis demonstrates that electronic polarization in BNNTs 
does not change its direction but rather grows monotonically with the
increasing diameter of the tube. Furthermore, the electronic and ionic spontaneous 
polarizations in BNNTs cancel exactly and these systems are pyroelectric only if 
their intrinsic helical symmetry is broken by, {\it e.g.,} inter-tube interactions 
or elastic distortions.

The rest of this paper is organized as follows: Sec.~\ref{sec_mtp}
briefly reviews the formulation of the modern polarization theory in terms of
Berry phases or Wannier functions. It also presents the details of the
numerical techniques that were used to compute polarization. In Sec.~\ref{sec_disc} 
we discuss the results and the complementary nature of the two techniques to
compute the spontaneous polarization. Finally, Sec.~\ref{sec_concl} presents
the summary and conclusions.

\section{computational methods}
\label{sec_mtp}

\subsection{Modern theory of polarization}

The problem of computing polarization in materials is very subtle and
is best approached by the ``Berry-phase'' method,
introduced only a decade ago by Vanderbilt and King-Smith,\cite{vanderbilt93}
and Resta.\cite{resta94}  Within this approach, the polarization
difference between two states of a system is computed as a
geometrical quantum phase.  In practice, this difference,
\hbox{$\Delta \bm{P} = \bm{P}^{(\lambda_1)} - \bm{P}^{(\lambda_0)}$},
can be obtained if one can find an adiabatic transformation $\lambda$
from one state to the other that leaves the system insulating. In
the spirit of Ref.~\onlinecite{vanderbilt00}, $\bm{P}^{(\lambda)}$ can
be split into two parts: $\bm{P}^{(\lambda)}_{ion}$ and
$\bm{P}^{(\lambda)}_{el}$, corresponding to the ionic and electronic
contributions respectively. In the case of paired
electron spins, the expression for the total polarization of the
system can be written as follows:
\begin{multline}
\bm{P}^{(\lambda)} = \bm{P}^{(\lambda)}_{ion} + \bm{P}^{(\lambda)}_{el}\\* 
= \frac{e}{V}\sum_{\tau} Z_{\tau}^{(\lambda)}
\bm{r}_{\tau}^{(\lambda)} - \frac{2 i e}{8 \pi^3} \sum_{i\, occ}
\int_{BZ} \!\!\!\! d \bm{k} \langle u_{i \bm{k}}^{(\lambda)} \big|
\nabla_{\bm{k}} \big| u_{i \bm{k}}^{(\lambda)} \rangle,
\label{polar_lambda}
\end{multline}
where $V$ is the volume of the unit cell, $Z_{\tau}$ and $\bm{r}_{\tau}$ 
are the charge and position of the $\tau$-th atom in the cell, and 
$u_{i \bm{k}}$ are the occupied cell-periodic Bloch states of the system.  
For the  \textit{electronic} part, an electronic phase
$\varphi_{\alpha}^{(\lambda)}$ (Berry phase) defined modulo $2 \pi$ can
be introduced as
\begin{equation}
\varphi_{\alpha}^{(\lambda)} = V \, \bm{G}_{\alpha} \cdot
\bm{P}_{el}^{(\lambda)} /e,
\label{berry_phase}
\end{equation}
where $\bm{G}_\alpha$ is the reciprocal lattice vector in the direction 
$\alpha$. Similarly, one can construct an angular variable for the
\textit{ionic} part, called in what follows the ``ionic'' phase,
so that the total geometrical phase is

\begin{equation}
\Phi_{\alpha}^{(\lambda)} = \sum_\tau Z_{\tau}^{(\lambda)} \bm{G}_{\alpha}
\cdot \bm{r}_{\tau}^{(\lambda)} + \varphi_{\alpha}^{(\lambda)}.
\label{total_phase}
\end{equation}
The total polarization in the direction $\alpha$ becomes
\begin{equation}
\bm{P}^{(\lambda)}_{\alpha} =
{e \Phi_{\alpha}^{(\lambda)} \bm{R}_{\alpha}}/{V},
\label{total_polar}
\end{equation}
where $\bm{R}_{\alpha}$ is the real-space lattice vector corresponding
to $\bm{G}_{\alpha}$, $(\bm{R}_{\alpha} \cdot \bm{G}_{\alpha}) = 1$.

Alternatively, the electronic polarization of a system can be
expressed in terms of the centers of charge of the
Wannier functions of its occupied bands:\cite{vanderbilt93,resta94}
\begin{equation}
\bm{P}^{(\lambda)}_{el} = -\frac{2e}{V}\sum_i
\int{\bf r}|W_i^{(\lambda)}({\bf r})|^2 d{\bf r} =
-\frac{2e}{V}\sum_i \langle {\bf r}^{(\lambda)}_i \rangle,
\label{polar_wannier}
\end{equation}
where the Wannier function (WF) $W_i^{(\lambda)}({\bf r})$ is constructed 
from the Bloch eigenstates $u_{i \bm{k}}^{(\lambda)}$  of band $i$ using the 
unitary transformation
\begin{equation}
W_i({\bf r}) = \frac{V}{(2 \pi)^3} \int_{BZ} \!\!\! e^{i \bm{k}\bm{r}} 
u_{i \bm{k}} ({\bf r}) d{\bf k},
\label{bstow}
\end{equation}
and $\langle {\bf r}^{(\lambda)}_i \rangle$ is the center of charge for the 
WF $W_i^{(\lambda)}$. However, 
because of the arbitrariness in the choice of the phases of the Bloch orbitals
(non-uniqueness of transformation~(\ref{bstow})), 
there is no unique representation of the WFs of a given group of bands. In our 
approach, we employ an algorithm that has been recently developed by Marzari and 
Vanderbilt,\cite{Marzari} which exploits the freedom in transformation~(\ref{bstow}) 
to construct WFs that are as localized as possible. This is achieved by minimizing 
the sum of the quadratic spreads of the Wannier probability 
distributions $|W_j ({\bf r})|^2$,
\begin{equation} 
\Omega=\sum_{j=1}^{M}[\langle r^2_j \rangle - \langle {\bf r}_j\rangle^2], 
\end{equation}
where the sum is over an isolated group of bands. The maximally localized 
WFs generated by this procedure are real, apart from an overall phase factor.

In both methods presented above, $\bm{P}_{el}^{(\lambda)}$ can be obtained 
only modulo $2e\bm{R}/V$ due to the arbitrariness in the choice of the phases 
of the Bloch functions. However, the difference in polarization $\Delta\bm{P}$ 
is well defined if $|\Delta \bm{P}_{el}| \ll |2e\bm{R}/V|$. The same 
indetermination issues apply to $\bm{P}_{ion}^{(\lambda)}$.\cite{vanderbilt00}

\subsection{Calculations}

In computing the spontaneous polarization as the difference
between a polar BNNT and a nonpolar reference state, a natural
choice for the nonpolar state is a nanotube of the same geometry,
but with boron and nitrogen atoms substituted by ``pseudo carbon''
atoms, which are 50\% boron and 50\% nitrogen. The adiabatic
transformation is then defined by a ``virtual crystal'' procedure,
in which parameter $\lambda$ corresponds to the content (in
atomic \%) of a site that is transformed from pure boron to the
nonpolar reference state (vice-versa for nitrogen sites). 

We used an \textit{ab initio} multigrid-based total-energy method, employing a 
real-space grid as a basis,\cite{briggs96} for all the Berry phase calculations
presented here. The Ce\-per\-ley-Al\-der \cite{ceperley80} form, parametrized by
Per\-dew and Zun\-ger,\cite{perdew81} was used for the exchange-correlation energy 
functional in the local density approximation. The norm-conserving pseudopotentials 
\cite{troullier91} for all the elements, including ``virtual'' ones,
were generated by the \texttt{fhi98PP} package\cite{fhi98pp} utilizing the 
Klein\-man-By\-lan\-der formulation.\cite{kleinman82}

To isolate the contribution of individual nanotubes, we performed polarization 
calculations for periodic crystals of \textit{noninteracting} (\textit{i.e.}, 
positioned sufficiently far apart) nanotubes in hexagonal and tetragonal
arrangements. The electronic structure
calculations were carried out using two special $k$-points along the
\hbox{$\Gamma$--$A$} direction in the hexagonal or \hbox{$\Gamma$--$Z$} direction  
in the tetragonal Brillouin zones. The $k$-space integration to compute 
$\varphi_z^{(\lambda)}$  was done on a string of 20 $k$-points uniformly distributed along
the same direction and shifted to avoid the $\Gamma$-point. The internal consistency 
of our approach was checked against the results obtained using the \texttt{ABINIT} 
code \cite{abinit}  for a few selected systems with excellent agreement. 

Because of the different alignments of the polar bond with respect to
the nanotube axis, we anticipate that the symmetry of the nanotube
will play an important role in determining the magnitude of the
spontaneous polarization field. In particular, since the zigzag
geometry maximizes the axial dipole moment, we expect to observe the
strongest effects in $(n,0)$ nanotubes.

\section{results and discussion}
\label{sec_disc}

\subsection{Berry-phase method}

\begin{figure}[ht!]
\includegraphics[scale=.475]{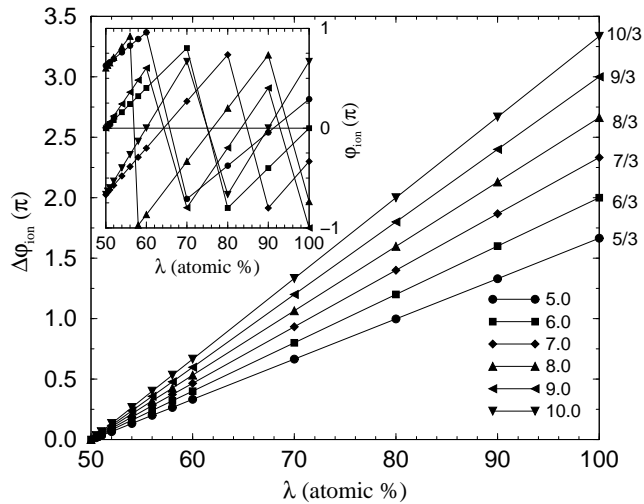}
\caption{ Ionic-phase difference between the polar and nonpolar
configurations for zigzag nanotubes; the ionic phase of
the nonpolar configuration is set to zero. Inset: ionic phases
wrapped into the \hbox{[$-\pi, \pi$]} interval. 
Phases are given in units of $\pi$.}
\label{ionic_phase_fig}
\end{figure}

The ionic part of the polarization in zigzag BNNTs, presented in 
Fig.~\ref{ionic_phase_fig}, is large and directly proportional to the 
nanotube's index. This is in contrast, for instance, to the
corresponding wurtzite III-V and II-VI systems,
\cite{zoroddu01,bernardini97} where the spontaneous polarization can
be viewed as the difference between the polarizations of the wurtzite
(polar) and zincblende (nonpolar) geometries. Since these
configurations become geometrically distinct only in the second shell
of neighbors, their ionic phases are very close.  The major
contribution to the spontaneous polarization in wurtzite materials is
then due to the difference between the electronic polarizations (which
are 0.04--0.08 C/m$^2$), while in BNNTs both the \textit{ionic} and
the \textit{electronic} contributions are essential. 

The ionic phase differences $\Delta\varphi_{ion}$ between the
polar and nonpolar configurations of zigzag nanotubes were
evaluated via the virtual crystal approximation. The inset in
Fig.~\ref{ionic_phase_fig} shows the results obtained by a simple
lattice summation over the ionic charges (the first term in 
Eq.~\ref{total_phase}), with the phases translated
into the \hbox{[$-\pi, \pi$]} interval. The phases plotted in the
main graph were ``unfolded'' by eliminating all the $2\pi$
discontinuities and setting the phase of the nonpolar reference
configuration to zero. For the unfolded phases, as the diameter of
a nanotube increases, \textit{i.e.,} as another hexagon is added
around the circumference of the tube, the ionic phase goes up by
$\pi/3$, so that the total ionic phase for a $(n,0)$ BN nanotube
amounts to $n\pi/3$. 

\begin{figure}[ht!]
\includegraphics[scale=.45]{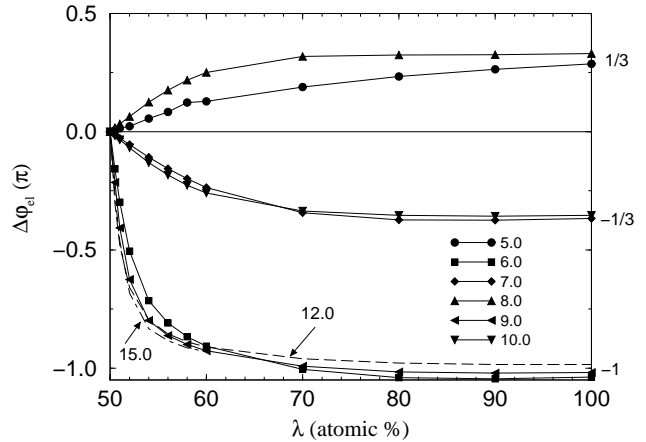}
\caption{ Electronic-phase differences between the polar and nonpolar
configurations for zigzag nanotubes.}
\label{electron_phase_fig}
\end{figure}

In Fig.~\ref{electron_phase_fig} we show the electronic-phase
differences $\Delta\varphi_{el}$ between the polar and nonpolar
configurations for zigzag nanotubes. These data suggest a natural
division of the nanotubes into three families with different
$\Delta\varphi_{el}$: $\pi/3$ for \hbox{$n = 3l - 1$,} $- \pi/3$ for
\hbox{$n = 3l + 1$,} and $- \pi$ for \hbox{$n = 3l$,} where $l$ is an
integer,\cite{flatsheet} which is similar to the result obtained by Mele and
Kr\'al.\cite{mele02} However, the existence of such three classes of
behavior is surprising, given that the ionic character of the
electronic charge density (associated with the B-N bond) does not
change with the nanotube index. Additionally, there is an important
difference between our results and those of
Ref.~\onlinecite{mele02}, where the electronic polarization of heteropolar
nanotubes was studied within a simple $\pi$-orbital tight-binding
($\pi$-TB) approximation. In Ref.~\onlinecite{mele02}, the
\hbox{``$n = 3l$''} family has a zero electronic phase instead of 
$-\pi$.

This discrepancy is due to the ambiguity of the definition of
electronic polarization as a multivalued quantity,
\cite{vanderbilt00} which can assume a lattice of values
corresponding to Berry phases that differ by arbitrary multiples
of $2\pi$. Unlike the ionic phase model, where discontinuities in
$\varphi_{ion}(\lambda)$ can be easily monitored, Berry phase
calculations always produce phases that are smoothly folded into
the $[-\pi,\pi]$ interval and cannot be extrapolated. To obtain an
unambiguous determination of the spontaneous polarization of BNNTs
of arbitrary diameters, one has to compute the polarization in a
different way, using the centers of charge of the WFs of the
occupied bands (Eq. \ref{polar_wannier}). Note that this approach
does not solve the problem of branch indetermination, since while
Berry phases are defined modulo $2\pi$, Wannier centers are
defined modulo a lattice vector {${\bf R}$. However, by shifting
the indetermination from the phase to the lattice vector, we are
able to map the electronic polarization problem onto a simple
electrostatic model, where the unfolding of the electronic phase
is straightforward.\cite{vanderbilt00}

\subsection{Maximally localized Wannier functions}

The results of the maximally localized WF calculations for BNNTs
are summarized in Fig.~\ref{wannier_fig}, where examples of the WFs 
for C and BN zigzag nanotubes of arbitrary diameter are shown, together
with a schematic drawing that illustrates the shift of the Wannier
centers in the adiabatic transformation from C to BN. Since
\begin{equation} 
\bm{P}^{(BN)}_{el} = -\frac{2e}{V} \sum_i({\bf r}_i^{(BN)} -
{\bf r}_i^{(C)}), 
\end{equation}
the magnitude of the shift of the centers is
directly proportional to the electronic polarization of the BNNT with
respect to the nonpolar CNT.

\begin{figure}[ht!]
\includegraphics[scale=.61]{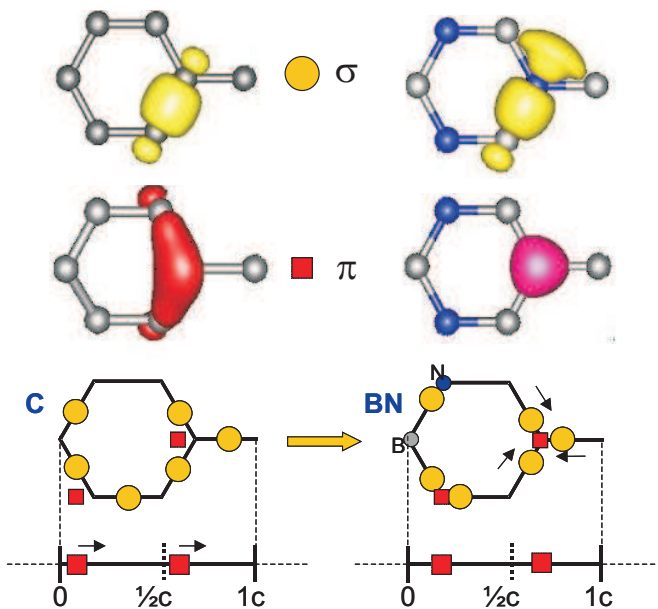}
\caption{ Upper panel: Examples of Wannier functions (WF) of the
$\sigma$ and $\pi$ occupied bands of C (left panel) and BN (right
panel) nanotubes. Lower panel: Schematic positions of the centers of
the Wannier functions in C and BN hexagons, and the projections of the
$\pi$ WF onto the nanotube axes. The positions of the centers of
$\sigma$ WF are indicated by circles, and those of $\pi$ by
squares. The direction of the shifts of $\sigma$ and $\pi$ WF in an
adiabatic transformation from C to BN is indicated by arrows.  The
projections of shifts of the $\sigma$ WF cancel, so that the $\sigma$
WF do not contribute to polarization (see text).}
\label{wannier_fig}
\end{figure}

The $\sigma$-band WFs are centered in the middle of the C-C bonds in
carbon nanotubes, while they are shifted towards the cations in BN
nanotubes because of the different electronegativities of B and N
atoms. Since these shifts have the same magnitude along each of the
three bond directions, the vector sum of all shifts is zero (see
bottom panel of Fig. \ref{wannier_fig}), and the $\sigma$ orbitals do
not contribute to the total polarization of the system. The $\pi$-band
WFs are centered on the cations in BNNTs, while in CNTs they have a
peculiar V-shape, with centers somewhat outside of the C-C bond.  The
sum of the shifts of the $\pi$-band Wannier centers is non-zero only
for the axial component, which means that the electronic polarization
in BNNTs is purely axial.

The bottom panel of Fig.~\ref{wannier_fig} shows the projection of the
$\pi$ WF centers onto the axis of the tube. The projections of the
centers have an effective periodicity of half of the axial lattice
constant $c$, which leads to the indetermination of the electronic
phase by multiples of $\pi$. Moreover, the WF
description allows for an unambiguous unfolding of the electronic
phase. In analogy to the ionic phase, we find that each individual
hexagon carries a phase of $-\pi/3$, leading to a total electronic
phase of $-n\pi/3$ for a $(n,0)$ nanotube. This result demonstrates
that the direction of the electronic polarization in a BNNT is
specified by the orientation of the B-N bond and does not oscillate in
direction with the nanotube diameter, contrary to the model
Hamiltonian predictions.\cite{mele02} We should point out that the
Wannier function results are completely consistent with the
Berry-phase calculations, since an electronic phase of $-n\pi/3$ for
any $n$ can be folded, modulo $\pi$, into the three families found
previously.

When we combine the results for the ionic and electronic
phases into a general formula for the phase of an arbitrary $(n,m)$ BNNT, 
\begin{multline}
\Delta \Phi_{z}^{tot} (n,m) = \Delta \varphi_{z}^{ion} (n,m) + 
\Delta \varphi_{z}^{el} (n,m) \\* 
= \frac{n-m}{3} - \frac{n-m}{3},
\end{multline}
we find that the two contributions cancel exactly and that
the total spontaneous polarization in any BNNT is zero, i.e.\ the
Wannier centers are arranged in such a way as to completely compensate
the polarization due to ions. We have verified
this result by two-point (\hbox{$\lambda=50$} and 100\%) calculations
of the Berry phase difference for a number of chiral nanotubes ((3,1),
(3,2), (4,1), (4,2), (5,2) and (8,2)) and found an exact cancellation
in all BNNTs, except for those narrower than approximately 4~\AA, where a residual
polarization is present as an effect of the very high curvature.
In such nanotubes Wannier centers cannot fully compensate the ionic polarization, 
due to the severe distortion of the atomic bonds, which makes these systems
weakly pyroelectric. For example, $P = 0.11$ C/m$^2$ in (3,1), 0.008 C/m$^2$ 
in (7,0) and 0.002 C/m$^2$ in (12,0) nanotubes. 

The exact cancellation is a result of the overall chiral symmetry of the
nanotubes which, although not centrosymmetric, are intrinsically nonpolar. 
Nevertheless, cancellation of ionic and electronic polarizations is
exact only in the limit of an isolated BNNT. The
spontaneous polarization in a nanotube bundle, where the chiral
symmetry is effectively broken, is different from zero. For example,
in (7,0) bundles at equilibrium distance of 3.2~\AA\ $P \approx 0.01$ C/m$^2$.
However, in this case it is hard to estimate the separate contributions to 
polarization due to bundling, extreme curvature and elastic deformation. 
Although smaller than in polymers or PZT, this polarization is comparable to 
some wurtzite pyroelectrics: {\it e.g.,} $P=0.06$ C/m$^2$ in 
{\it w}-ZnO.\cite{bernardini97}

\subsection{Piezoelectricity}

\begin{table}[t!]
\tabcolsep=0.08cm
\renewcommand{\arraystretch}{1.1}

\caption{Piezoelectric properties of zigzag BNNT
bundles.\cite{NT_crystal} The corresponding values for a few
piezoelectric materials are listed for comparison.}
\label{polar_data}
\begin{ruledtabular}
\begin{tabular}{ccccc}
$(n,m)$ & diameter (\AA) & $Z^*$ (e) & $|e_{33}|$ (C/m$^2$) & Ref.\\
%$(n,m)$ & diameter & $Z^*$  & $e_{31}$  & $|e_{33}|$ \\
%$     $ & (\AA)    &  (e)   & (C/m$^2$) & (C/m$^2$)  \\

\hline
\ph{1}(5,0) & 3.91 &  2.739 & 0.389 & \\
\ph{1}(6,0) & 4.69 &  2.696 & 0.332 & \\
\ph{1}(7,0) & 5.47 &  2.655 & 0.293 & \\
\ph{1}(8,0) & 6.24 &  2.639 & 0.263 & \\
\ph{1}(9,0) & 7.04 &  2.634 & 0.239 & \\
     (10,0) & 7.83 &  2.626 & 0.224 & \\
     (11,0) & 8.57 &  2.614 & 0.211 & \\
     (12,0) & 9.38 &  2.609 & 0.198 & \\     
     (13,0) &10.16 &  2.605 & 0.186 & \\
\hline
% \textit{w}-AlN\footnotemark[1]
%      \footnotetext[1]{Ref. \onlinecite{zoroddu01}.}    &  & 2.653 & -0.53  & 1.50 \\
% \textit{w}-ZnO\footnotemark[2]							   
%      \footnotetext[2]{Ref. \onlinecite{bernardini97}.} &  & 2.11  & -0.51  & 0.89 \\
% PbTiO$_3$\footnotemark[3]							   
%      \footnotetext[3]{Ref. \onlinecite{PT}.}           &  &       & -0.93  & 3.23 \\
% P(VDF/TrFE)\footnotemark[4]
%      \footnotetext[4]{PVDF copolymer with trifluoroethylene (TrFe). Ref. \onlinecite{PVDF}.}

\textit{w}-AlN &  & 2.653 & 1.50 & [\onlinecite{zoroddu01}]\\
\textit{w}-ZnO &  & 2.11  & 0.89 & [\onlinecite{bernardini97}] \\
    PbTiO$_3$  &  &       & 3.23 & [\onlinecite{PT}]\\
P(VDF/TrFE)    &  &       & $\approx$ 0.12 & [\onlinecite{PVDF}] \\
\end{tabular}
\end{ruledtabular}
\end{table}

The Berry-phase method can also be employed to compute
piezoelectric properties of BNNTs, which are directly related to
polarization differences between strained and unstrained
tubes. In the linear regime, the change in
polarization due to strain can be decomposed into a sum of two terms:
a uniform axial strain and a relative displacement of the two
sublattices. It is therefore natural to describe the geometry of a
BNNT of a given radius in terms of an axial lattice constant $c$
and an internal parameter $u$, where $uc$ is the length of the vector
connecting the anion with the cation. With this choice, the axial 
piezoelectric polarization is
\begin{equation}
\delta P_3 = e_{33} \epsilon_3 =
\frac{\partial P_3}{\partial c} (c - c_0) +
\frac{\partial P_3}{\partial u} (u - u_0),
\label{piezo_deriv}
\end{equation}
where the strain is $\epsilon_3 = (c - c_0)/c_0$, and $c_0$ and $u_0$ are
the equilibrium values of $c$ and $u$. The only surviving
piezoelectric strain tensor component is 
\begin{equation}
e_{33} = e_{33}^{(0)} + \frac{e c_0^2}{V} N Z^* \frac{du}{dc},
\label{piezo_const}
\end{equation}
where $N$ is the number of B-N pairs in the supercell. Here, 
\begin{equation}
e_{33}^{(0)} = c_0 {\partial P_3}/{\partial c} 
\end{equation}
is the ``clamped-ion'' piezoelectric constant (representing the 
effect of strain on the electronic structure), and 
\begin{equation}
Z^* = ({V}/{e N c_0}){\partial P_3}/{\partial u}
\end{equation}
is the axial component of the Born dynamical charge
tensor. Both polarization derivatives were computed as
finite differences, changing $c$ or $u$ by $\pm 1$\%.
The parameter $\xi = c_0 du/dc$, 
describing the change in the bond lengths under axial strain,
was obtained by rescaling $c$ together with the 
associated components of ionic coordinates, and then relaxing the
geometry of the system. For all the systems considered below, the value
of $\xi$ is approximately the same and equal to -0.085.

We have calculated the piezoelectric properties for various bundles
comprised of zigzag BNNTs with individual diameters ranging from 3.9
to 10.2~\AA. These results are summarized in Table~\ref{polar_data} and
compared to a few well-known piezo- and pyroelectric materials.
While the piezoelectric constants of zigzag BNNTs are modest when
compared with inorganic compounds, they are still substantially larger
than those in the PVDF polymer family.

\section{summary and conclusions}
\label{sec_concl}

In summary, we have investigated the spontaneous polarization and
piezoelectric properties of BN nanotubes using state-of-the-art
\textit{ab initio} methods. Our calculations demonstrate the
complementary nature of Berry phase and Wannier function analysis, and
show that a real-space description is necessary to unravel the Berry
phases in complicated cases. The results suggest that BNNTs are excellent
nonpolar piezoelectrics that exhibit substantially higher strain response 
than polar polymers. Moreover, we have shown that, contrary to the previous
expectations, ideal non-interacting nanotubes are effectively
nonpolar due to their intrinsic chiral symmetry, which leads to a total
cancellation between the ionic and electronic polarizations. Breakage of this
symmetry, as in the simple case of interacting nanotubes in a bundle,
induces spontaneous polarization fields that are comparable to
those of wurtzite semiconductors. Due to their piezo- and pyroelectric
properties, BNNTs are excellent
candidates for various nano-electro-mechanical applications.

\section*{Acknowledgments}

We are indebted to N. Marzari and I. Souza for their invaluable
help in computing Wannier functions in nanotubes.  We would also like
to thank D. Vanderbilt and R. Resta for illuminating discussions. This
work was supported by NASA, ONR and the Mathematical, Information
and Computational Sciences Division, Office of Advanced Scientific
Computing Reasearch of the U.S.\ D.O.E.\ under Contract
No.\ DE-AC05-00OR22725 with UT-Battelle, LLC.
The calculations were carried out at DoD and NC Supercomputing Centers.


\begin{thebibliography}{99}
\bibitem{flatsheet} In the limiting case of flat C or BN sheets, the
electronic polarization is zero for any value of $\lambda$, due to the
existence of a threefold symmetry axis perpendicular to the surface of
the sheet.
\end{thebibliography}
\end{document}